\documentstyle[preprint,prd,aps]{revtex}
\advance\textheight by \baselineskip
\begin{document}
\draft

\preprintstytrue
\preprint{\hbox to\textwidth{UCLA/95/TEP/21 \hfill
    hep-th/9506085 \hfill UCONN-95-3}}

\title{Effective Energy for QED$_{2+1}$ \\
 with Semi-Localized Static Magnetic Fields: A Solvable Model
  }

\author{Daniel Cangemi\cite{emaila} and Eric D'Hoker\cite{emailb}}

\address{ Department of Physics, University of California Los Angeles,
  Los Angeles, CA 90095-1547 } \author{Gerald Dunne\cite{emailc}}

\address{ Department of Physics, University of Connecticut, Storrs, CT
  06269-3046 }

\date{June 1995}

\maketitle

\begin{abstract}%
  We evaluate the exact ${\rm QED}_{2+1}$ effective energy for charged
  spin zero and spin half fields in the presence of a family of static
  magnetic field profiles localized in a strip of width $\lambda$. The
  exact result yields an infinite set of relations between the terms
  in the derivative expansion of the effective energy for a general magnetic
  field. Upon addition of the standard Maxwell magneto-static energy, the
  minimum energy configuration at fixed flux corresponds to a uniform magnetic
  field.
\end{abstract}


The effective action and the effective energy are important tools for
the study of quantum electrodynamics \cite{schwinger,das1}. Using the
proper-time technique, Schwinger showed \cite{schwinger} that the QED
effective action can be computed exactly for either a {\it constant}
or a {\it plane wave} electro-magnetic field. This result was later
adapted to ${\rm QED}_{2+1}$ with constant fields by Redlich
\cite{redlich}. For more general electro-magnetic fields one generally
performs some sort of perturbative expansion, such as the derivative
expansion \cite{aitchison} which yields the large distance behavior of
the theory. This approach was recently applied to ${\rm QED}_{2+1}$
\cite{cangemi}, showing (for example) for the special case of zero
electric field and static magnetic field, that while the zero
derivative term increases the effective energy, the next order
correction term with two derivatives tends to decrease the effective
energy. The question of vacuum stability is inaccessible in a
derivative expansion, so more powerful tools are required to study it.
In this Paper we make a first step towards the non-perturbative
understanding of such a system by considering a new exactly solvable
model which has a spatially-varying magnetic field.  We expect our
model to be relevant for recent investigations of symmetry breaking
\cite{hosotani,miransky} and finite temperature effects
\cite{andersen,das2} in ${\rm QED}_{2+1}$.

We show that the ${\rm QED}_{2+1}$ effective energy
for charged spin zero and spin half particles of arbitrary mass can be
computed {\it exactly} in the presence of time-independent but
spatially-varying magnetic fields of the form:
\begin{equation}
  B(x,y) = \frac{B}{[\cosh (x/\lambda) ]^2}
\label{mag}
\end{equation}
Here, the magnetic field is localized in a strip of infinite
extent in the $y$-direction and of width $\lambda$ in the
$x$-direction. In the limit $\lambda \to \infty$ the magnetic field in
(\ref{mag}) tends to a uniform one of strength $B$. The constant strength $B$
sets a length scale $1/\sqrt{eB}$ known as the magnetic length, and the
derivative expansion regime corresponds to $\lambda>>1/\sqrt{eB}$. In practice,
the system will be considered in a box in the $y$-direction of size $L$. The
total flux $\Phi$ of the magnetic
field is then finite : $\Phi = eB\lambda L/\pi$. We show that the
effective energy has a simple {\it exact} integral representation
involving elementary functions for all values of mass $m$, width
$\lambda$, electric charge $e$, and strength $B$.

The motivation for the choice of these particular profiles for the
magnetic field resides in the fact that they may be fairly
representative of the type of inhomogeneity that we expect in the
system, while still being exactly solvable. They have been chosen here in the
form of string-like flux tubes with finite width and infinite length. While we
cannot, of course, analyze exactly the dynamics for arbitrary magnetic field
configurations, the exact solution will permit us to study the system
non-perturbatively within this family of profiles and gain rigorous constraints
on the derivative expansion for more general fields.

The starting point for the evaluation of the effective energy in
the presence of the magnetic field (1) is the functional determinant
expression of the effective action, which we quote in Minkowski
space-time :
\begin{equation}
  i\int d^3x \,{\cal L} _\pm = \mp \ln ~\text{Det} ~\{D_\mu D^\mu +
  m^2 + e \Sigma _\pm ^{\mu \nu} F_{\mu \nu} -i \epsilon \}.
\end{equation}
Here, ${\cal L}_\pm$ are the effective Lagrangians for bosons ($+$)
and fermions ($-$) of electric charge $e$ in the presence of a gauge
potential $A_\mu$ with $D_\mu = \partial _\mu + i e A_\mu$. The
Lagrangian ${\cal L}_-$ with $\Sigma _- ^{\mu \nu} = (i/4) [ \gamma
^\mu,\gamma ^\nu]$ produces the effective action for a 4-component
spinor consisting of 2-component spinors of masses $m$ and $-m$
respectively, whereas ${\cal L}_+$ with $\Sigma _+ ^{\mu\nu}=0$
produces the effective action for spin 0 complex scalars with mass
$m$. Both systems are invariant under parity and time reversal. The
determinants are understood to be regulated by Pauli-Villars masses in
the ultra-violet, which will not be exhibited explicitly. Also, we
calculate the effective action relative to that for zero
electro-magnetic fields and therefore drop all contributions
independent of the fields. The effective energy for static magnetic
fields is then given by ${\cal E}_\pm = -\int dx~dy~{\cal L}_\pm$.

For the family of magnetic fields in (1), we may use the translation
invariance of the problem in time and in the $y$-direction to work in
an eigenbasis of frequency $\omega$ and $y$-momentum $k$. Then the
operator $ D_\mu D^\mu + m^2 +e \Sigma _\pm ^{\mu \nu} F_{\mu \nu}$
coincides with the Schr\"odinger operator of a solvable 1-dimensional
quantum mechanical system. (Recall that the solvability of the
constant field case is based on its relation to the solvable 1-dimensional
harmonic oscillator system \cite{weisskopf}). To see this, we choose
the gauge potential $A_x=0$ and $A_y = \lambda B \tanh (x/\lambda)$,
which reproduces the magnetic field in (1).
\begin{mathletters}
\begin{eqnarray}
  D_\mu D^\mu + m^2 + e \Sigma _\pm ^{\mu \nu} F_{\mu \nu} =
  -\frac{d^2}{dx^2} + V_k(x) - \omega ^2
\label{schrodinger}
\end{eqnarray}
\begin{eqnarray}
  V_k(x) = - \case{1}{\lambda ^2}(\gamma_\pm^2 - \case{1}{4}) [1 -
  (\tanh\case{x}{\lambda})^2] + \case{1}{2} \alpha_k^2 (1 +
  \tanh\case{x}{\lambda}) + \case{1}{2} \alpha_{-k}^2 (1 -
  \tanh\case{x}{\lambda})
\end{eqnarray}
\end{mathletters}%
Here, the following assignments for the parameters of this potential
have been made, with $\sigma ^3$ denoting the spin projection
eigenvalue $+1$ for spin up fermions and $-1$ for spin down fermions:
\begin{mathletters}
\begin{eqnarray}
  &\alpha_k = \sqrt{(k - eB \lambda)^2 + m^2} \\[6pt] &\gamma_\pm =
  \left\{\begin{array}{ll} \case{1}{2} \sqrt{1 + 4 ( eB\lambda^2)^2} &
      (+) \text{ bosons} \\[5pt] \case{1}{2} + eB\lambda^2 \sigma ^3 &
      (-) \text{ fermions }
  \end{array} \right.
\end{eqnarray}
\end{mathletters}%
The 1-dimensional Schr\"odinger operator (\ref{schrodinger}) is of the
supersymmetric quantum mechanical form \cite{cooper}, but with a
$k$-dependent super-potential $W_k(x)=\lambda B \tanh (x/\lambda)-k$.
The $k$-dependent potential $V_k(x)$ is of the modified
P\"oschl-Teller form, with asymptotic energy barrier heights
$\alpha_{\pm k}^2$ as $x \to \pm \infty$; the corresponding incoming
and outgoing momenta are then just $\alpha _{\pm k}$, while $\omega
^2$ is merely an overall shift in energy. The origin of the difference
between the two asymptotic energy barriers may be simply understood in
terms of classical electrodynamics. As a particle of charge $e$ enters
the magnetic strip from $-\infty$, its kinetic energy is conserved
while its momentum in the $y$ direction behaves as $p_y(x) =
p_y(-\infty) +eB \lambda (\tanh \case {x}{\lambda} +1)$. If the
momentum in the $x$ direction is large enough, the particle traverses
the strip, but for small $x$-momentum, it will instead be reflected
off the strip and return to $-\infty$.

The evaluation of the determinants in (2) thus reduces to a spectral
problem for the Schr\"odinger operator in (\ref{schrodinger}). Note
that this Schr\"odinger operator has both a discrete and a continuous
spectrum, in contrast to the constant B field case for which the
spectrum is purely discrete. It is convenient at this point to
analytically continue frequencies to imaginary values : $\omega \to i
\omega$, as usual; the effective energy is then given by
\begin{equation}
  {\cal E}_\pm = \pm \frac{L}{4 \pi ^2} \int _{-\infty} ^{+\infty} dk
  \int_{-\infty} ^{+\infty} d\omega ~ \ln \text{Det} \left(-
    \frac{d^2}{dx^2} + V_k(x) + \omega ^2 \right)
\end{equation}
[Notice that we actually compute the difference between ${\cal E}_\pm$ and the
corresponding free field ($B=0$) case; thus we may drop terms independent of
$B$.] Integrating by parts in $\omega$ and omitting $B$-independent terms, the
effective energy may be recast as
\begin{equation}
  {\cal E} _\pm = \mp \frac{2 L}{4\pi ^2} \int _{-\infty} ^{+\infty}
  dk \int _{-\infty} ^{+\infty} d\omega ~\omega ^2~\text{Tr} ~
  G_{-\omega ^2,k}
\label{effective}
\end{equation}
The resolvent Green function $G_{E,k}$ is the inverse of the
Schr\"odinger operator, for general complex parameter $E$ :
\begin{equation}
  \left(- \frac{d^2}{dx^2} + V_k(x) - E \right) G_{E,k}(x, x') =
  \delta ( x - x')
\label{green}
\end{equation}
It is standard to obtain $G_{E,k}(x,x')$ by matching the independent
solutions of the homogeneous equation, which are proportional to
hyper-geometric functions in terms of the new variable $ \xi =[1 +\tanh
(x/\lambda)]/2$ :
\begin{mathletters}
\begin{eqnarray}
  & u_1(x) = \xi^\alpha (1-\xi)^\beta F(z+\case{1}{2} -\gamma _\pm
,z+\case{1}{2}
  +\gamma _\pm;1+2\alpha;\xi) \\[6pt] & u_2(x) = \xi ^\alpha
  (1-\xi)^\beta F(z+\case{1}{2} - \gamma _\pm ,z+\case{1}{2} +\gamma
  _\pm;1+2\beta;1-\xi)
\end{eqnarray}
\end{mathletters}%
Here, $\alpha $ and $\beta$ are defined as the roots with positive
real part of the following equations.
\begin{equation}
  \alpha = \case{\lambda }{2}\sqrt{\alpha_{-k}^2 - E} \hspace{34pt}
  \beta = \case{\lambda }{2}\sqrt{\alpha _{k}^2 - E} \hspace{34pt}
  z=\alpha +\beta
\end{equation}
With these conventions, $u_1$ is regular at $\xi=0$ (i.e. as $x \to
-\infty$), whereas $u_2$ is regular at $\xi=1$ (i.e. as $x \to
+\infty$).  The Wronskian of these solutions is a constant, given by
\begin{equation}
  W = u_1'(x) u_2(x) - u_1(x) u'_2(x) = \frac{2}{\lambda}
  \frac{\Gamma(1+2\alpha) \Gamma(1+2\beta)}{\Gamma(z+\case{1}{2} -\gamma
    _\pm) \Gamma(z+\case{1}{2} +\gamma _\pm)}
\end{equation}
and the Green function $G_{E,k}$ is therefore given by
\begin{equation}
  G_{E,k}(x,x') = \theta(x'-x) \frac{u_1(x) u_2(x')}{W} +\theta(x-x')
  \frac{u_1(x') u_2(x)}{W}
\end{equation}
All information concerning the spectrum of the Schr\"odinger operator
is contained in the trace of the Green function $G_{E,k}$. We first
compute
\begin{equation}
  \text{Tr} ~G_{E,k} = \int _{-\infty} ^ {+\infty} dx ~G_{E,k} (x,x) =
  \frac{1}{W}\int _{-\infty} ^ {+\infty} dx ~u_1(x) u_2(x)
\end{equation}
Some care is needed in regularizing this integral at $x=\pm \infty$;
for example, one can multiply the arguments $\xi$ and $1-\xi$ in the
hyper-geometric functions in $u_1$ and $u_2$ respectively by a factor
$1-\epsilon$ for $\epsilon >0$ and infinitesimal.  With this
regularization, the integral may be performed exactly in terms of the
Euler psi-function $\psi(x)= \Gamma '(x) /\Gamma(x)$, and we find
\begin{equation}
  \text{Tr} ~G_{E,k} = - \case{\lambda^2}{4} \left(\frac{1}{\alpha} +
    \frac{1}{\beta} \right) \left[ \psi(z+\case{1}{2} -\gamma _\pm) +
    \psi(z+\case{1}{2} +\gamma _\pm) \right] + f_\epsilon(\alpha) +
  f_\epsilon(\beta)
\label{trace}
\end{equation}
where $f_\epsilon(\alpha)$ is a function of $\alpha$, but not of
$\beta$, whose precise form is regulator dependent, but which does not
contribute to the effective energy once we integrate over $k$. Here and in
the following, it
is understood that for the case of 4-spinors $(-)$ both spin states
are to be summed over. The spectrum contains a finite number of bound
states, which arise from the (simple) poles of the $\psi$-functions in
(\ref{trace}) at $z+\case{1}{2} -\gamma_\pm=-n$ for $0\leq n < \gamma_\pm -
1/2 - \sqrt{|\alpha_k^2 - \alpha_{-k}^2|}$:
\begin{equation}
  E_n = \case{1}{2} (\alpha_k^2 + \alpha_{-k}^2) - \lambda ^{-2}(n +
  \case{1}{2} -\gamma_\pm)^2 - \case{1}{16} \lambda^2 (\alpha_k^2 -
  \alpha_{-k}^2)^2 (n + \case{1}{2} -\gamma_\pm)^{-2}
\end{equation}
The same discrete spectrum may be obtained by solving the homogeneous
Schr\"odinger equation (\ref{green}) for real $E$ and requiring
normalizability of the eigenfunctions \cite{morse}. The spectrum also
contains a cut starting at $\alpha_k^2$ and another cut starting at
$\alpha _{-k}^2$, corresponding to the two barrier thresholds. For
$B=0$, the discrete spectrum is absent, whereas for constant magnetic
field case ({\it i.e.} $\lambda \to \infty$), it reduces to $E_n = 2
eB (n+ 1/2)$ for bosons and $2 eB n$ for fermions, as expected.

We now complete the calculation of the effective energy, using the
result of (\ref{trace}) in the expression (\ref{effective}) for the
effective energy.  First, since $\alpha$ only depends upon $k+
eB\lambda$, we may shift $k$ by $-eB\lambda$ in the contribution of
$f_\epsilon(\alpha)$ in (\ref{trace}). Thus, the regulator-dependent
$f_\epsilon$ terms in (\ref{trace}) yield only $B$-independent
contributions to the effective energy in (\ref{effective}) and may be
omitted.  Next, we use the identity
\begin{equation}
  -\frac{\lambda^2}{8} \left(\frac{1}{\alpha} + \frac{1}{\beta}
  \right) = \frac{\partial z}{\partial E}
\end{equation}
which suggests the change of variable from $\omega$ to
$z=\alpha+\beta$ and yields
\begin{equation}
  {\cal E}_\pm = \pm \frac{L}{2\pi ^2 \lambda }
  \int_{-\infty}^{+\infty} \!\!\! dk \int_{|\alpha_k + \alpha_{-k}|}
  ^\infty \!\!\!\! dz\,\frac{1}{z} \sqrt{(z^2 - (\alpha_k +
    \alpha_{-k}) ^2) (z^2 - (\alpha_k - \alpha_{-k})^2)} \bigl [\psi(z
  + \case{1}{2} - \gamma_\pm) + \psi(z + \case{1}{2} + \gamma_\pm) \bigr ]
\end{equation}
The integration over $k$ can be performed and we end up with the
remarkably simple expression
\begin{equation}
  {\cal E}_\pm = \pm \frac{L}{4\pi\lambda^2} \int_{z_0}^\infty dz\;
  \frac{z (z^2 - z_0^2)}{\sqrt{z^2 - z_0^2 + (\lambda m)^2}} \; \bigl
  [ \psi(z + \case{1}{2} - \gamma_\pm) + \psi(z + \case{1}{2} + \gamma_\pm)
\bigr
  ]
\label{simple}
\end{equation}
where $ z_0 \equiv \lambda \sqrt{(eB\lambda )^2 + m^2}$. This
expression can be rewritten in terms of elementary integrals by making
use of the following representation of the $\psi$-function
\cite{bateman}:
\begin{equation}
  \psi (x) = \ln x -\frac{ 1}{2x} -2 \int _0 ^\infty
  \frac{t~dt}{(t^2+x^2)(e^{2\pi t} -1)}
\label{psifunction}
\end{equation}
The first two terms on the right hand side in (\ref{psifunction})
contribute $B$-independent terms to ${\cal E}_\pm$. (It is necessary
to sum over both spin states to see this in the fermion case.) For
the third term, the $z$-integral can be carried out exactly and we
obtain the following finite integral representation for the effective
energy
\begin{equation}
  {\cal E}_\pm(L,m\lambda,eB\lambda^2) = \frac{L}{4\pi\lambda^2} \int_
  0 ^\infty dt ~\frac{1}{e^{2 \pi t}\pm 1} ~\left ( (b_\pm-it)
    \frac{(\lambda ^2 m^2+v_\pm^2)}{v_\pm} \ln \frac{\lambda m
      -iv_\pm}{\lambda m +iv_\pm} + c.c. \right )
\label{answer}
\end{equation}
where $c.c.$ denotes the complex conjugate, and
\begin{mathletters}
\begin{equation}
  b_\pm=\left\{\begin{array}{ll} \sqrt{(eB\lambda^2)^2+1/4} & \;\;\;
      (+) \text{\ \ bosons} \\ eB\lambda^2 & \;\;\; (-) \text{\ \
        fermions} \end{array} \right.
\end{equation}
\begin{equation}
  v_\pm^2=\left\{\begin{array}{ll} t^2 + 2i\,t\,b_+-1/4 & \;\;\; (+)
      \text{\ \ bosons}\\ t^2 + 2i\,t\,b_- & \;\;\; (-) \text{\ \
        fermions} \end{array} \right.
\end{equation}
\end{mathletters}%
Expression~(\ref{answer}) gives the exact effective action for a
background field (\ref{mag}) and is the main result of this
Paper. For definiteness, we now concentrate on the
fermion case, but analogous discussions can be made for bosons.

In the limit of vanishing mass, $m=0$, the only relevant dimensionless
parameter is $eB\lambda^2$, and so one finds an asymptotic expansion
\begin{eqnarray}
  {\cal E}_- &=& \frac{L \lambda(eB)^{3/2}}{8 \pi}\sum_{j=0}^\infty
  \frac{1}{(4\pi eB\lambda^2)^j} \frac{\Gamma(j-3/2)\Gamma(j+5/2)}{
    \Gamma(j+1) \Gamma(j/2+1/4) \Gamma(3/4-j/2)} \zeta(j+3/2) \\ &=&
  \frac{L \lambda(eB)^{3/2}}{8\sqrt{2} \pi}\left[ \zeta(3/2)-
    \frac{15}{ 16\pi}\zeta(5/2) \frac{1}{eB\lambda^2}+\dots\right]
\end{eqnarray}
where $\zeta(z)$ is the Riemann zeta function \cite{bateman}. The first term in
this expansion agrees with the uniform $B$ field case \cite{redlich,blau},
while the next term agrees with the first-order derivative expansion
computation in \cite{cangemi}.

For nonzero mass $m$, there is another dimensionless parameter,
$eB/m^2$, which is the ratio of the cyclotron energy to the rest mass
energy. A double expansion of (\ref{answer}) yields
\begin{equation}
  {\cal E}_- =  \frac{L m^3 \lambda}{8 \pi} \sum _{j=0}^\infty
  \frac{1}{j!} (2eB\lambda^2)^{(-j)} \sum_{k=1}^\infty
  \frac{(2k+j-1)!}{(2 k)!} \frac{{\cal B}_{2k+2j}}{(2k+j-1/2)(2k+j-3/2)}
  \left(\frac{2eB}{m^2} \right)^{2k+j}
\label{doublesum}
\end{equation}
with ${\cal B}_{n}$ the $n^{th}$ Bernoulli number \cite{bateman}. Each
power in $\lambda^{-2}$ corresponds to a fixed order in a derivative
expansion of the effective action.  The zeroth order term agrees with
the $eB/m^2$ expansion of the exact constant $B$ field answer
\cite{redlich,blau}, while the first order term agrees with the
$eB/m^2$ expansion of the leading derivative expansion contribution
found in \cite{cangemi}.

In fact, the specific configurations~(\ref{mag}) give some insight in
the more general case of a background magnetic field $B(x)$ depending
on one coordinate only. In a derivative expansion of the effective
Lagrangian the terms with a total of $2j$ derivatives, ${\cal
  L}^{[2j]}$, have, up to integration by parts, a unique structure
\begin{mathletters}
\begin{eqnarray}
  {\cal L}^{[0]} &=& m^3 \, F^{[0]}_0(\case{eB(x)}{m^2}) \\
  {\cal L}^{[2]} &=& m \, F^{[2]}_2(\case{eB(x)}{m^2}) \, \left[\frac{e
    B'(x)}{m^2}\right]^2 \\
  {\cal L}^{[2j]} &=&
    m^{3-2j} \sum_{l=1}^{2j-2} F^{[2j]}_l(\case{eB(x)}{m^2}) \,
  \frac{eB^{(2j-l)}(x)}{m^2} \, \left[\frac{eB'(x)}{m^2}\right]^{l}
  \qquad 2j=4,6,8,\ldots
\end{eqnarray}
\end{mathletters}%
where $B^{(l)}(x)$ denotes the $l$-th derivative of $B(x)$. Parity
invariance forces $F^{[2j]}_l(x)$ to be even (resp. odd) for $l$ odd
(resp. even). A comparison with (\ref{doublesum}) entirely determines
the zero- and two-derivative terms (in agreement with~\cite{cangemi})
and gives at each order ($2j>2$) a relation among the $(2j-2)$
functions $F^{[2j]}_l(x)$.
\begin{mathletters}
\label{recur}
\begin{eqnarray}
\lefteqn{ F^{[0]}_0(x) = - \sum_{k=1}^\infty \frac{1}{8\pi^{3/2}} \,
  \frac{\Gamma(2k-3/2)}{\Gamma(2k+1)} \, {\cal B}_{2k} \; (2x)^{2k}} \\
\lefteqn{ F^{[2]}_2(x) = - \sum_{k=1}^\infty \frac{1}{4\pi^{3/2}} \,
  \frac{\Gamma(2k-1/2)}{\Gamma(2k)} \, {\cal B}_{2k+2} \; (2x)^{2k-2}} \\
\lefteqn{ \sum_{l=1}^{j-1} \sum_{s=1}^{j-l+1} \, (-2)^{2j-s} \,
  \frac{\Gamma(j+3/2-s)}{\Gamma(3/2)} \, \left(
    \frac{d}{dx} \right)^{s-1} x^{2l+s-2} \Biggl\{ \, \left[s
    \, a^s_{j-l+1} + 2 (j-l+2-s) \, a^{s-1}_{j-l+1} \right] \;
    F^{[2j]}_{2l-1}(x)} \hspace{1.5in} \nonumber \\
  &&{}+ x \, a^s_{j-l+1} \; F^{[2j]}_{2l}(x) \, \Biggr\} =
  \sum_{k=1}^\infty
    \frac{1}{8\pi^{3/2}} \frac{\Gamma(2k+j)}{\Gamma(2k+1)}
    \frac{\Gamma(2k+j-3/2)}{\Gamma(2k)} \, {\cal B}_{2k+2j} \; (2 x)^{2k-1}
\end{eqnarray}
\end{mathletters}%
where $a^s_p$ are the coefficients in the polynomial of degree $2p$
\begin{equation}
  \left[(1-t^2)\frac{d}{dt}\right]^{2p-1} \, t \equiv \sum_{s=1}^p
  a^s_p \, (-2)^{2p-s-1} \, t^{2p-2s} \, (1-t^2)^s
\end{equation}
As an example, the two functions appearing in the four-derivative
terms ($2j=4$) obey
\begin{equation}
  \left(4 \frac{d}{dx} + \frac{5}{x}\right) \, F^{[4]}_1(x) + x\frac{d}{dx}
    \, F^{[4]}_2(x) = \sum_{k=1}^\infty \frac{1}{8\pi^{3/2}} (2k+1)
    \frac{\Gamma(2k+1/2)}{\Gamma(2k)} \, {\cal B}_{2k+4} \; (2
    x)^{2k-3}
\label{recur4}
\end{equation}
Thus, Eqs. (\ref{recur}) provide an infinite number of relations among
the coefficients of the effective action for any background magnetic
field with translation invariance in one direction.

Finally, we may apply the results on the effective energy obtained
here and in~\cite{cangemi} to the study of the full QED$_{2+1}$
theory.  In~\cite{cangemi}, the effective energy is obtained for large
wavelength fluctuations in the fields. While the leading order term
contributes positively to the energy, the next, two-derivative term
contributes negatively. It was proposed in~\cite{cangemi} that this
behavior may drive the system towards a lowest energy state with
inhomogeneous magnetic field. Indeed, since flux is conserved in
QED$_{2+1}$, it is natural to consider static fluctuations $\delta
B(\vec x)$ that leave the total flux unchanged. Under this constraint,
the energy fluctuation is given by
\begin{equation}
  {\cal E}(B+\delta B(\vec{x})) - {\cal E}(B) = \frac{1}{2}\int d^2x
  \, \left\{1 + \alpha \, (e^2/\sqrt{eB}) \left[1 - \beta \, (eB)^{-1}
    |\vec{\partial} \delta B/\delta B|^2 \right]
  \right\} (\delta B)^2
\label{delB}
\end{equation}
where the coefficients $\alpha$ and $\beta$ are positive functions of
$m^2/(eB)$, except for bosons with sufficiently small mass as
explained in~\cite{cangemi}. The coefficient $\beta$ is typically of order $1$,
and within the derivative expansion regime $(eB)^{-1}|\vec{\partial} \delta
B/\delta B|^2<<1$, indicating that the constant background magnetic field is
stable under variations that conserve flux.

In the case of the special family of profiles for the magnetic field
in~(\ref{mag}) we can go beyond the leading order derivative expansion, using
the exact effective energy (\ref{answer}), together with the Maxwell term, to
give the total energy as
\begin{equation}
  {\cal E}^{\text{tot}}_\pm = L \lambda B^2 + {\cal E}_\pm(L, m
  \lambda,eB \lambda^2) =
  \frac{\pi^2}{e^2} \frac{\Phi^2}{\lambda L} + {\cal
    E}_\pm(L, m\lambda,\pi\Phi\lambda/L)
\label{totenergy}
\end{equation}
with flux $\Phi=eB\lambda L/\pi$. We have shown that the total
energy~(\ref{totenergy}) at fixed flux $\Phi$ is a positive
monotonically decreasing function of the parameter $\lambda$. As a
result, a system with magnetic profile~(\ref{mag}) is driven towards a
system with uniform magnetic field.  Thus, within this family, we have
not found further support for our suggestion in~\cite{cangemi} that
magnetic field inhomogeneities lower the total energy at fixed
non-zero flux.

This raises a number of open questions. (a) Whether any other families
of magnetic field profiles are integrable, within which non-uniform
magnetic fields minimize the energy at fixed flux. The recursion
relations~(\ref{recur}) suggest that magnetic field profiles with
$x$-dependence only may be such candidates. (b) More generally,
whether inhomogeneous magnetic fields can be found that minimize the
energy for fixed flux.

\bigskip
This work is supported in part by the NSF under contract
PHY-92-18990 and by the DOE under grant DE-FG02-92ER40716.00.

\end{document}